\ifx\pdfoutput\undefined
	\documentclass[dvips,apjl]{emulateapj}
  \DeclareGraphicsExtensions{.eps}
\else
	\documentclass[pdftex,apjl]{emulateapj}
  \DeclareGraphicsExtensions{.pdf}
\fi

\usepackage{amsmath}
\usepackage{apjfonts}

\begin{document}

\title{On the exact analytic expressions for the equitemporal surfaces in Gamma-Ray Burst afterglows.}

\author{Carlo Luciano Bianco\altaffilmark{1} and Remo Ruffini\altaffilmark{2}}
\affil{ICRA --- International Center for Relativistic Astrophysics.}
\affil{Dipartimento di Fisica, Universit\`a di Roma ``La Sapienza'', Piazzale Aldo Moro 5, I-00185 Roma, Italy.}

\altaffiltext{1}{E-mail: bianco@icra.it}
\altaffiltext{2}{E-mail: ruffini@icra.it}

\begin{abstract}
We have recently shown \citep[see][]{EQTS_ApJL} that marked differences exist between the EQuiTemporal Surfaces (EQTSs) for the Gamma-Ray Burst (GRB) afterglows numerically computed by the full integration of the equations of motion and the ones found in the current literature expressed analytically on the grounds of various approximations. In this Letter the exact analytic expressions of the EQTSs are presented both in the case of fully radiative and adiabatic regimes. The new EQTS analytic solutions validate the numerical results obtained in \citet{EQTS_ApJL} and offer a powerful tool to analytically perform the estimates of the physical observables in GRB afterglows.
\end{abstract}


\section{Introduction}

We have recently shown \citep[see][]{EQTS_ApJL} that in Gamma-Ray Bursts (GRBs) marked differences exist between the equitemporal surfaces (EQTSs) for the afterglow numerically computed by the full integration of the equations of motion and the ones found in the current literature, expressed analytically on the grounds of various approximations. Indeed, the approximate formulae in the current literature overestimate the size of the EQTSs approximately by a factor of 27\% or 20\% in the early part of the afterglow ($t_a^d=35$ s), in the adiabatic and fully radiative cases respectively. Correspondingly, they underestimate the size of the EQTSs approximately by a factor of 15\% or 28\% in the latest part of the afterglow ($t_a^d=4$ days). The precise knowledge of the EQTSs is essential to obtain the observational properties of GRBs.

In this Letter, progress is made in making manifest the difference between the exact expressions of the EQTSs and the ones obtained by approximate methods in the current literature: the exact analytic expressions of the EQTSs are found in the case of both fully radiative and adiabatic regimes.

\section{The EQTS for GRBs}

\subsection{The afterglow hydrodynamical equations}

The discovery of the afterglow \citep{c97} has offered a very powerful tool for the understanding of GRBs. Consensus has been reached that such an afterglow originates from a relativistic shock front propagating in the InterStellar Medium (ISM) and that its description can be obtained from energy and momentum conservation in relativistic hydrodynamics. Consensus exists, as well, that the shock fluid is concentrated in a thin shell. The fulfillment of the energy and momentum conservation in the laboratory reference frame leads to \citep[see e.g.][and references therein]{p99,Brasile}:
\begin{equation}
\left\{\begin{aligned}
dE_{\mathrm{int}} & = \left(\gamma - 1\right) dM_{\mathrm{ism}} c^2\\
d\gamma & = - \frac{{\gamma}^2 - 1}{M} dM_{\mathrm{ism}}\\
dM & = \frac{1-\varepsilon}{c^2}dE_{\mathrm{int}}+dM_\mathrm{ism}\\
dM_\mathrm{ism} & = 4\pi m_p n_\mathrm{ism} r^2 dr \end{aligned}\right.
\label{Taub_Eq}
\end{equation}
where $\gamma$, $E_\mathrm{int}$ and $M$ are the pulse Lorentz gamma factor, internal energy and mass-energy respectively, $n_\mathrm{ism}$ is the ISM number density, $m_p$ is the proton mass, $\varepsilon$ is the emitted fraction of the energy developed in the collision with the ISM and $M_\mathrm{ism}$ is the amount of ISM mass swept up within the radius $r$:
\begin{equation}
M_\mathrm{ism}=m_pn_\mathrm{ism}\frac{4\pi}{3}\left(r^3-{r_\circ}^3\right)\, ,
\label{dgm1}
\end{equation}
where $r_\circ$ is the starting radius of the shock front. In general, an additional equation is needed in order to express the dependence of $\varepsilon$ on the radial coordinate. In the following, $\varepsilon$ is assumed to be constant and such an approximation appears to be correct in the GRB context.

\subsection{The definition of the EQTSs}

For the case of a spherically symmetric expansion considered in this Letter, the EQTSs are surfaces of revolution about the line of sight. The general expression for their profile, in the form $\vartheta = \vartheta(r)$, corresponding to an arrival time $t_a$ of the photons at the detector, can be obtained from \citep[see e.g.][]{Brasile,EQTS_ApJL}:
\begin{equation}
ct_a = ct\left(r\right) - r\cos \vartheta  + r^\star\, ,
\label{ta_g}
\end{equation}
where $r^\star$ is the initial size of the expanding source, $\vartheta$ is the angle between the radial expansion velocity of a point on its surface and the line of sight, and $t = t(r)$ is its equation of motion, expressed in the laboratory frame, obtained by the integration of Eqs.(\ref{Taub_Eq}). From the definition of the Lorentz gamma factor $\gamma^{-2}=1-(dr/cdt)^2$, we have in fact:
\begin{equation}
ct\left(r\right)=\int_0^r\left[1-\gamma^{-2}\left(r'\right)\right]^{-1/2}dr'\, ,
\label{tdir}
\end{equation}
where $\gamma(r)$ comes from the integration of Eqs.(\ref{Taub_Eq}).

It is appropriate to underline a basic difference between the apparent superluminal velocity orthogonal to the line of sight, $v^\bot \simeq \gamma v$, and the apparent superluminal velocity along the line of sight, $v^\parallel \simeq \gamma^2 v$. In the case of GRBs, this last one is the most relevant: for a Lorentz gamma factor $\gamma \simeq 300$ we have $v^\parallel \simeq 10^5 c$. This is self-consistently verified in the structure of the ``prompt radiation'' of GRBs, see e.g. \citet{rbcfx02_letter}.

\subsection{The case of an adiabatic regime}

We first examine the case of an adiabatic regime ($\varepsilon = 0$). The dynamics of the system is described by the following solution of the Eqs.(\ref{Taub_Eq}) \citep[see e.g.][and references therein]{p99}:
\begin{equation}
\gamma^2=\frac{\gamma_\circ^2+2\gamma_\circ\left(M_\mathrm{ism}/M_B\right)+\left(M_\mathrm{ism}/M_B\right)^2}{1+2\gamma_\circ\left(M_\mathrm{ism}/M_B\right)+\left(M_\mathrm{ism}/M_B\right)^2}\, ,
\label{gamma_ad}
\end{equation}
where $\gamma_\circ$ and $M_B$ are respectively the values of the Lorentz gamma factor and of the mass of the accelerated baryons at the beginning of the afterglow phase.

We have performed an exact analytic integration of Eq.(\ref{tdir}) using Eq.(\ref{gamma_ad}) and, as a consequence, we have the exact analytic solution:
\begin{equation}
t\left(r\right) = \left(\gamma_\circ-\frac{m_i^\circ}{M_B}\right)\frac{r-r_\circ}{c\sqrt{\gamma_\circ^2-1}} + \frac{m_i^\circ}{4M_Br_\circ^3}\frac{r^4-r_\circ^4}{c\sqrt{\gamma_\circ^2-1}} + t_\circ\, ,
\label{analsol_ad}
\end{equation}
where $t_\circ$ is the value of the time $t$ at the beginning of the afterglow phase and $m_i^\circ=(4/3)\pi m_p n_{\mathrm{ism}} r_\circ^3$.

The analytic expression for the EQTS in the adiabatic regime can then be obtained substituting $t(r)$ from Eq.(\ref{analsol_ad}) in Eq.(\ref{ta_g}). We obtain:
\begin{equation}
\begin{split}
\cos\vartheta & = \frac{m_i^\circ}{4M_B\sqrt{\gamma_\circ^2-1}}\left[\left(\frac{r}{r_\circ}\right)^3  - \frac{r_\circ}{r}\right] + \frac{ct_\circ}{r} \\[6pt] & - \frac{ct_a}{r} + \frac{r^\star}{r} - \frac{\gamma_\circ-\left(m_i^\circ/M_B\right)}{\sqrt{\gamma_\circ^2-1}}\left[\frac{r_\circ}{r} - 1\right]\, .
\end{split}
\label{eqts_g_dopo_ad}
\end{equation}

\subsection{The case of a fully radiative regime}

We turn now to the case of a fully radiative regime ($\varepsilon = 1$). The dynamics of the system is given by the following solution of the Eqs.(\ref{Taub_Eq}) \citep[see e.g.][and references therein]{p99,Brasile}:
\begin{equation}
\gamma=\frac{1+\left(M_\mathrm{ism}/M_B\right)\left(1+\gamma_\circ^{-1}\right)\left[1+\left(1/2\right)\left(M_\mathrm{ism}/M_B\right)\right]}{\gamma_\circ^{-1}+\left(M_\mathrm{ism}/M_B\right)\left(1+\gamma_\circ^{-1}\right)\left[1+\left(1/2\right)\left(M_\mathrm{ism}/M_B\right)\right]}\, .
\label{gamma_rad}
\end{equation}

Again, like in the adiabatic case, we have performed an exact analytic integration of Eq.(\ref{tdir}) using Eq.(\ref{gamma_rad}). As a consequence, we have \citep{Brasile}:
\begin{equation}
\begin{split}
t\left(r\right) & = \frac{M_B  - m_i^\circ}{2c\sqrt C }\left( {r - r_\circ } \right) + \frac{m_i^\circ r_\circ }{8c\sqrt C }\left[ {\left( {\frac{r}{{r_\circ }}} \right)^4  - 1} \right] \\[6pt]
& + \frac{{r_\circ \sqrt C }}{{12cm_i^\circ A^2 }} \ln \left\{ {\frac{{\left[ {A + \left(r/r_\circ\right)} \right]^3 \left(A^3  + 1\right)}}{{\left[A^3  + \left( r/r_\circ \right)^3\right] \left( {A + 1} \right)^3}}} \right\} + t_\circ \\[6pt]
& + \frac{{r_\circ \sqrt{3C}}}{{6 c m_i^\circ A^2 }} \left[\arctan \frac{{2\left(r/r_\circ\right) - A}}{{A\sqrt 3 }} - \arctan \frac{{2 - A}}{{A\sqrt 3 }}\right]\, ,
\end{split}
\label{analsol}
\end{equation}
where $A=\sqrt[3]{\left(M_B-m_i^\circ\right)/m_i^\circ}$ and $C={M_B}^2(\gamma_\circ-1)/(\gamma_\circ +1)$.

The analytic expression for the EQTS in the fully radiative regime can then be obtained substituting $t(r)$ from Eq.(\ref{analsol}) in Eq.(\ref{ta_g}). We obtain:
\begin{equation}
\begin{split}
&\cos\vartheta=\frac{M_B  - m_i^\circ}{2r\sqrt{C}}\left( {r - r_\circ } \right) +\frac{m_i^\circ r_\circ }{8r\sqrt{C}}\left[ {\left( {\frac{r}{{r_\circ }}} \right)^4  - 1} \right] \\[6pt]
&+\frac{{r_\circ \sqrt{C} }}{{12rm_i^\circ A^2 }} \ln \left\{ {\frac{{\left[ {A + \left(r/r_\circ\right)} \right]^3 \left(A^3  + 1\right)}}{{\left[A^3  + \left( r/r_\circ \right)^3\right] \left( {A + 1} \right)^3}}} \right\} +\frac{ct_\circ}{r}-\frac{ct_a}{r} \\[6pt] & + \frac{r^\star}{r} +\frac{{r_\circ \sqrt{3C} }}{{6rm_i^\circ A^2 }} \left[ \arctan \frac{{2\left(r/r_\circ\right) - A}}{{A\sqrt{3} }} - \arctan \frac{{2 - A}}{{A\sqrt{3} }}\right]\, .
\end{split}
\label{eqts_g_dopo}
\end{equation}

\subsection{Comparison between the two cases}

\begin{figure}
\includegraphics[width=\hsize,clip]{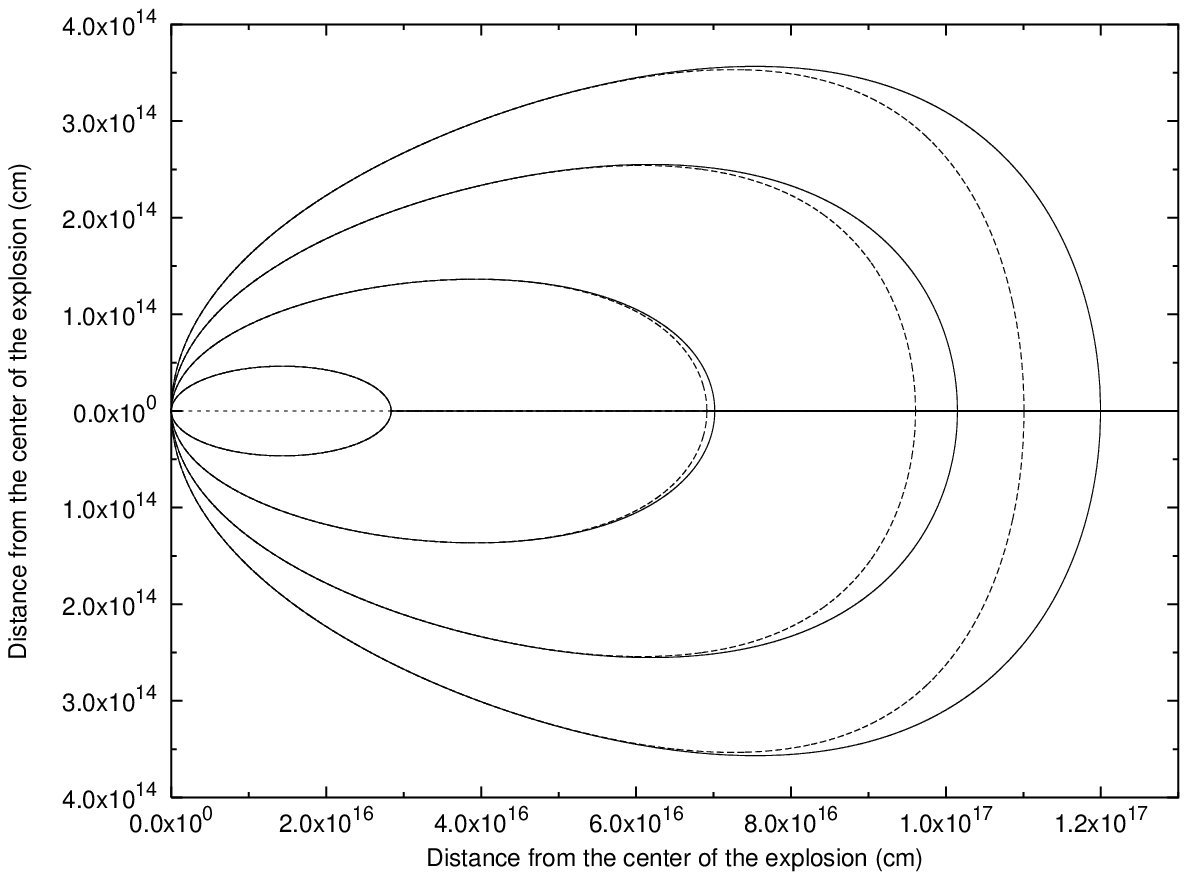}
\includegraphics[width=\hsize,clip]{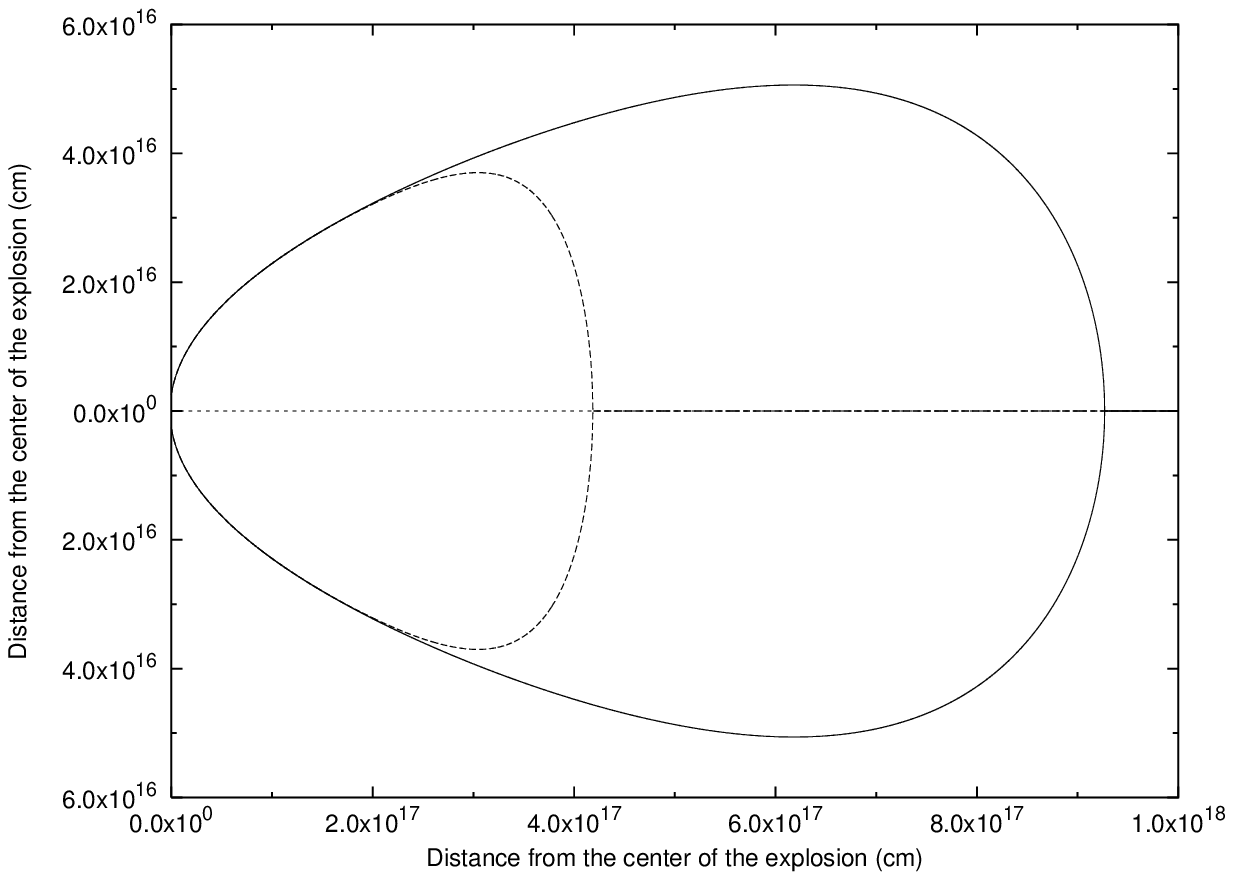}
\caption{Comparison between EQTSs in the adiabatic regime (solid lines) and in the fully radiative regime (dashed lines). The upper plot shows the EQTSs for $t_a=5$ s, $t_a=15$ s, $t_a=30$ s and $t_a=45$ s, respectively from the inner to the outer one. The lower plot shows the EQTS at an arrival time of 2 days.}
\label{eqts_comp}
\end{figure}

The two EQTSs are represented at selected values of the arrival time $t_a$ in Fig.~\ref{eqts_comp}, where the illustrative case of GRB~991216 has been used as a prototype. The initial conditions at the beginning of the afterglow era are in this case given by $\gamma_\circ = 310.131$, $r_\circ = 1.943 \times 10^{14}$ cm, $t_\circ = 6.481 \times 10^{3}$ s, $r^\star = 2.354 \times 10^8$ cm \citep[see][]{lett1,lett2,rbcfx02_letter,Brasile}.

\section{Approximations adopted in the current literature}\label{approx}

In the current literature two different treatments of the EQTSs exist: one by \citet{pm98c} and one by \citet{s98} later applied also by \citet{gps99} \citep[see also][and references therein]{p99,p00,vpkw00}.

In both these treatments, instead of the more precise dynamical equations given in Eqs.(\ref{gamma_ad},\ref{gamma_rad}), the following simplified formula, based on the ``ultrarelativistic'' approximation, has been used:
\begin{equation}
\gamma \propto r^{-\alpha}\, ,
\label{gamma_app}
\end{equation}
where $\alpha = 3$ in the fully radiative and $\alpha = 3/2$ in the adiabatic cases. A critical analysis comparing and contrasting our exact solutions with Eq.(\ref{gamma_app}) has been presented in \citet{Power_Laws}. As a further approximation, instead of the exact Eq.(\ref{tdir}), they both use the following expansion at first order in $\gamma^{-2}$:
\begin{equation}
ct\left(r\right) = \int_0^r \left[1+\frac{1}{2\gamma^2\left(r'\right)}\right] dr'\, .
\label{tdir_app}
\end{equation}
Correspondingly, instead of the exact Eq.(\ref{analsol_ad}) and Eq.(\ref{analsol}), they find:
\begin{subequations}\label{t_app_pm98c_s98}
\begin{eqnarray}
t\left(r\right) & = & \frac{r}{c}\left[1+\frac{1}{2\left(2\alpha+1\right)\gamma^2\left(r\right)}\right]\, ,\\
t\left(r\right) & = & \frac{r}{c}\left[1+\frac{1}{16\gamma^2\left(r\right)}\right]\, .
\end{eqnarray}
\end{subequations}
The first expression has been given by \citet{pm98c} and applies both in the adiabatic ($\alpha=3/2$) and in the fully radiative ($\alpha=3$) cases (see their Eq.(2)). The second one has been given by \citet{s98} in the adiabatic case (see his Eq.(2)). Note that the first expression, in the case $\alpha=3/2$, does not coincide with the second one: \citet{s98} uses a Lorentz gamma factor $\Gamma$ of a shock front propagating in the expanding pulse, with $\Gamma = \sqrt{2} \gamma$. Without entering into the relative merit of such differing approaches, we show that both of them lead to results very different from our exact solutions.

Instead of the exact Eqs.(\ref{ta_g}), \citet{pm98c} and \citet{s98} both uses the following equation:
\begin{equation}
ct_a = ct\left(r\right) - r\cos \vartheta\, ,
\label{ta_g_app}
\end{equation}
where the initial size $r^\star$ has been neglected. The following approximate expressions for the EQTSs have been then presented:
\begin{subequations}\label{eqts_app}
\begin{eqnarray}
\vartheta & = & 2\arcsin\left[\frac{1}{2\gamma_\circ}\sqrt{\frac{2\gamma_\circ^2ct_a}{r}-\frac{1}{2\alpha+1}\left(\frac{r}{r_\circ}\right)^{2\alpha}}\right]\, ,\\
\cos\vartheta & = & 1-\frac{1}{16\gamma_L^2}\left[\left(\frac{r}{r_L}\right)^{-1}-\left(\frac{r}{r_L}\right)^{3}\right]\, .
\end{eqnarray}
\end{subequations}
The first expression has been given by \citet{pm98c} and applies both in the adiabatic ($\alpha=3/2$) and in the fully radiative ($\alpha=3$) cases (see their Eq.(3)). The second expression, where $\gamma_L \equiv \gamma(\vartheta=0)$ over the given EQTS and $r_L=16\gamma_L^2ct_a$, has been given by \citet{s98} in the adiabatic case (see his Eq.(5)).

In \citet{EQTS_ApJL} we have compared and contrasted the results of the approximate expressions given in Eqs.(\ref{eqts_app}) with the ones based on the exact solutions, there numerically computed and here given for the first time in analytic form in Eqs.(\ref{eqts_g_dopo_ad},\ref{eqts_g_dopo}).

\section{Conclusions}

The formulae we have obtained are manifestly different from the ones in the current literature. They are valid for any value of the Lorentz gamma factor and they may be applied, as well, to the physics and astrophysics of supernovae and active galactic nuclei. However, as suggested by the referee, a word of caution is appropriate: the applicability of the thin shell approximation, used in deriving Eqs.\eqref{Taub_Eq}, is likely to break down when the non relativistic Newtonian phase is approached. There, the swept up ISM mass is no longer concentrated in a thin shell as exemplified, e.g., by the Sedov-Taylor-Von Neumann solution \citep[see e.g.][]{sedov}.

The new EQTS analytic solutions validate the numerical results obtained in \citet{EQTS_ApJL}. We have indeed verified the perfect agreement between the results of the numerical computations, presented there, and the new analytic results, presented here.

From the numerical examples given in \citet{EQTS_ApJL} it is also clear that differences exist between the correct treatment and the approximate ones all along the GRB afterglow process: the approximate treatments systematically overestimate the size of the EQTSs in the early part of the afterglow and underestimate it in the latest part.

The analytic results presented in this Letter, when applied to a specific model of the shock front emission process \citep{Spectr1} duly taking into account the ISM filamentary structure \citep{Spectr2}, allows to make precise predictions of the luminosity in fixed energy bands and of the instantaneous as well as time integrated spectra of GRB afterglow.

\acknowledgements

We are thankful to the anonymous referee for very good suggestions on the manuscript.

\end{document}